# High Spectral Energy Density All-Fiber Nanosecond Pulsed 1.7 μm Light Source for Photoacoustic Microscopy

Seongjin Bak[a,1], Sang Min Park[a,1], Yuon Song[b], Jeesu Kim[a,b*], Tae Won Nam[c], Dong-Wook Han[b], Chang-Seok Kim[a,b], Soon-Woo Cho[d*], Brett E. Bouma[e], Hwidon Lee[a,b*]

[a]*Engineering Research Center for Color-Modulated Extra-Sensory Perception Technology, Pusan National University, Busan, 46241, Republic of Korea*

[b]*Department of Cogno-Mechatronics Engineering, Pusan National University, Busan 46241, Republic of Korea*

[c]*Department of Materials and Science and Engineering, Pusan National University, Busan, 46241, Republic of Korea*

[d]*Department of Biomedical Engineering, Duke University, Durham, NC, 27708, USA*

[e]*Wellman Center for Photomedicine, Harvard Medical School and Massachusetts General Hospital, 40 Blossom Street, Boston, MA, 02114, USA*

[1]*These authors contributed equally to this work*

**Corresponding author: jeesukim@pusan.ac.kr, soon-woo.cho@duke.edu and hwidonlee@pusan.ac.kr*

**Abstract**

We present a high spectral energy density all-fiber nanosecond pulsed 1.7 μm light source specifically designed for photoacoustic microscopy (PAM). The system targets the 1st overtone absorption of C–H bonds near 1720 nm within the near-infrared-III (NIR-III) window, where lipids exhibit strong optical absorption, and tissues benefit from reduced scattering and high permissible fluence. To achieve narrow-linewidth, high pulse energy, and high pulse repetition rate (PRR), we developed a master oscillator fiber amplifier architecture based on stimulated Raman scattering. A 1589.80 nm Raman pump and a custom-built narrow-linewidth Raman seed laser were employed to generate spectrally pure 1719.44 nm pulses (~0.10 nm linewidth). The proposed light source delivers nanosecond pulses (~5 ns) with high pulse energy (≥2.2 μJ) and tunable PRRs up to 300 kHz, resulting in a spectral energy density of approximately 22 μJ/nm—significantly higher than that of conventional 1.7 μm light sources. Performance of the NIR-PAM system was validated through resolution testing with a 1951 USAF target, demonstrating a spatial resolution of approximately 4.14 μm and an axial resolution of approximately 85.5 μm. Phantom imaging of $CH_2$-rich polymer films and ex vivo lipid-rich biological tissues confirmed the system's high spatial fidelity and strong contrast for lipid-specific structures. This compact, stable, and spectrally refined light source with high spectral energy density can offer an effective solution for high-resolution, label-free molecular imaging and represents a promising platform for clinical photoacoustic imaging applications involving lipid detection and metabolic disease diagnostics.

# 1. Introduction

Photoacoustic imaging (PAI) is an advanced biomedical imaging modality that synergistically integrates functional optical imaging with the deep tissue penetration capabilities of ultrasound. PAI relies on the generation of ultrasound waves (so-called photoacoustic (PA) waves) induced by pulsed laser irradiation of endogenous tissue chromophores. These chromophores absorb optical energy at specific wavelengths corresponding to their intrinsic molecular vibrations, thereby enabling the extraction of molecular and functional information without the need for exogenous contrast agents [1–12]. Consequently, the careful selection of the light source's wavelength that corresponds to the absorption peaks of target chromophores is crucial for maximizing imaging contrast and overall performance.

Hemoglobin has been extensively studied as a primary chromophore in PAI due to the availability of high-power excitation at 532 nm — readily obtained via second harmonic generation from a Nd:YAG laser operating at 1064 nm. While hemoglobin imaging has laid a strong foundation for PAI, fully harnessing its potential to resolve optical absorption with high contrast necessitates expanding the scope to other biologically relevant chromophores, such as melanin [13,14], collagen [15–17] and lipids [18–20]. Consequently, a variety of light sources have been developed and optimized across the near-infrared (NIR) spectral region to effectively visualize these chromophores in biological tissues [21–24].

The NIR-III optical window (1600–1870 nm) has especially attracted considerable attention in biomedical imaging owing to its reduced optical scattering and minimized local effects, which collectively enable high-contrast, deep-tissue imaging. Specific wavelengths within this spectral region not only support low-phototoxicity imaging across a variety of tissue types [25], but also coincide with the absorption bands of lipids—key biomolecules that play critical roles in energy storage, membrane composition, and biosynthesis [26]. Given these biological functions, dysregulation of lipid metabolism has been strongly linked to a range of pathological conditions, including metabolic disorders, diabetes, and hepatic dysfunction [27,28]. Furthermore, excessive lipid accumulation is a key contributor to the onset and progression of cardiovascular diseases, such as atherosclerosis, one of the leading causes of mortality worldwide [29,30]. Therefore, the development of light sources operating at optimal wavelengths for effective lipid detection is of crucial for advancing PAI toward the early diagnosis and treatment of lipid-related diseases.

The absorption spectrum of lipids, predominantly by C–H bond vibrations, exhibits multiple overlapping peaks corresponding to specific vibrational modes [31]. Within the NIR region (700–2500 nm), two prominent absorption bands are observed: the 2$^{nd}$ overtone around 1210 nm and the 1$^{st}$ overtone near 1720 nm [31,32]. Among these, the 1700 nm spectral region has gained particular attention for deep-tissue imaging of lipids, not only due to reduced scattering and lower phototoxicity but also because the 1$^{st}$ overtone exhibits a vibrational transition strength approximately an order of magnitude greater than that of the 2nd overtone. Correspondingly, PA signals generated at ~1700 nm have been reported to be approximately 6.3 times stronger than those at ~1200 nm [32,33]. Moreover, the maximum permissible exposure (MPE) in the 1700 nm region, as defined by the American National Standards Institute, is 1 J/cm$^2$— 10 times higher than the MPE in the 1200 nm region [34] —thereby allowing for higher excitation energy while maintaining safety. This increased MPE translates into enhanced PA signal generation and improved signal-to-noise ratio (SNR) at deeper imaging depths [33]. Taken together, the excitation near 1700 nm offers distinct advantages over shorter wavelengths for achieving high SNR and deep-tissue imaging of lipids with minimal phototoxic effect [35,36].

To exploit the strong lipid absorption near the 1700 nm spectral region, various light sources have been employed in PAI, including optical parametric oscillators (OPOs), supercontinuum (SC) lasers, gain-switched thulium-doped fiber (TDF) lasers, and stimulated Raman scattering (SRS) lasers —each offering distinct advantages and facing specific limitations [3,19,21,23,37,38]. Crucial performance parameters such as output wavelength, pulse width, pulse repetition rate (PRR), and pulse energy collectively determine key aspects of PAI, including chemical bond selectivity, system resolution, A-scan acquisition speed, and molecular detection sensitivity. In particular, spectral energy density plays a critical role in enabling selective excitation of molecular vibrations, thereby enhancing chemical bond specificity [39]. OPOs, for instance, allow for versatile wavelength tuning via nonlinear optical conversion; however, they necessitate precise phase matching, which leads to increased system complexity, higher costs, and typically limited PRRs in the kilohertz range [3,37]. SC lasers provide broadband spectral coverage but often suffer from low spectral energy density, typically below 100 nJ/nm [23,38]. Gain-switched TDF lasers have also been explored; however, their relatively long pulse width (~16 ns) and a fixed low PRR (~10 kHz) restrict their utility for high-speed, high-resolution imaging [19]. More recently, synchronously pumped Raman fiber laser (SPRFL) has demonstrated improvements over conventional light sources. Nonetheless, it still encounters critical challenges, including a fixed and wavelength-dependent PRR (~100 kHz), limited spectral energy density (<500 nJ/nm), and increased system complexity due to the need for precise synchronization between the pulsed Raman pump and the external Raman laser cavity. Additionally, achieving high PRRs over 100 kHz requires a long optical fiber cavity longer than 500 m, which introduces temperature-sensitive instabilities that may degrade overall system performance [21].

For effective PAI, particularly in high-resolution photoacoustic microscopy (PAM), the excitation light source must satisfy several critical requirements: stable output power to ensure consistent PA signal generation; tunable, high PRR to support diverse system configurations and imaging depths; and a narrow spectral linewidth centered at the molecular absorption peak to enable high SNR and spectral specificity [40–42]. Additionally, high spectral energy density is crucial to efficiently excite specific vibrational modes, especially when targeting overtone absorption bands such as those of C–H bonds.

To address these requirements, we developed a high spectral energy density all-fiber nanosecond pulsed 1.7 μm light source specifically optimized for lipid-selective PAM in the NIR-III window. The system is based on a master oscillator fiber amplifier (MOFA) architecture utilizing the SRS effect, providing flexible control of PRR. The output wavelength of 1719.44 nm was carefully chosen to match the 1st overtone absorption peak of lipids, while the Raman pump at 1589.80 nm delivers an optimal Raman frequency shift (~14.3 THz) for efficient L-band amplification [43]. A custom-built narrow-linewidth continuous-wave (CW) Raman seed laser was developed to ensure high spectral energy density SRS output.

The resulting system provides high pulse energy (≥2.2 μJ), narrow-linewidth (~0.10 nm), short pulse duration (~5 ns), and tunable PRR up to 300 kHz. The combination of high pulse energy and narrow-linewidth yields a high spectral energy density of approximately 22 μJ/nm. These results indicate the feasibility of future integration into high-speed and multimodal PAI platforms, with the potential for enhanced SNR through signal averaging [44]. System performance was validated through phantom and tissue imaging, demonstrating high-resolution, high-contrast visualization of lipid structures. This compact, stable light source represents a promising platform for label-free molecular imaging and clinical applications in lipid-related disease diagnostics.

# 2. Materials and methods

## *2.1. System configuration*

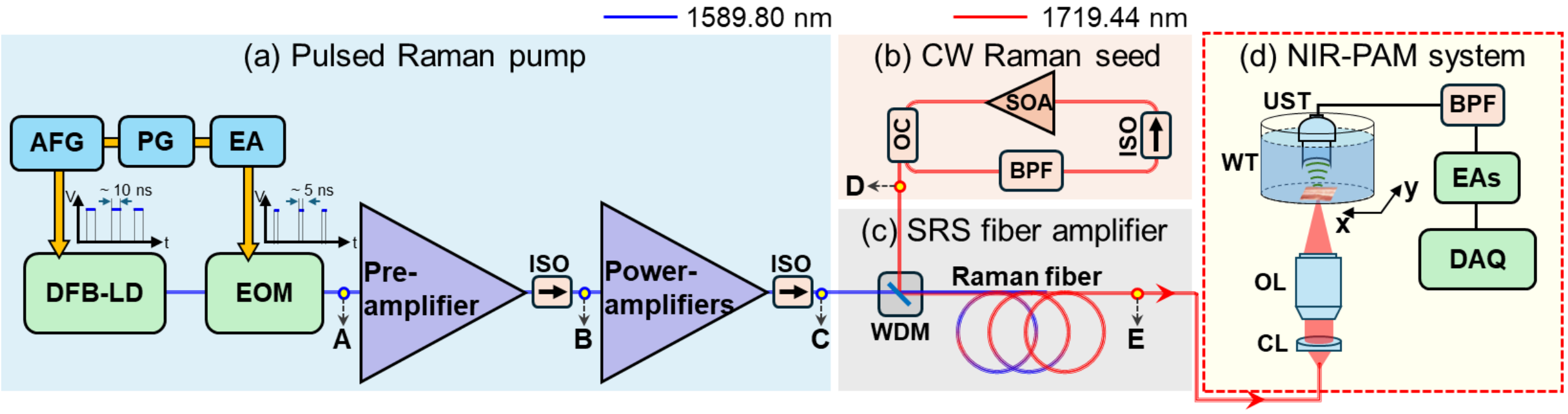

**Fig. 1** Schematic of the high spectral energy density all-fiber nanosecond pulsed 1.7 μm stimulated Raman scattering (SRS) amplifier and near-infrared photoacoustic microscopy (NIR-PAM) system. (a) Pulsed Raman pump: DFB-LD (distributed-feedback laser diode), EOM (electro-optic modulator), AFG (arbitrary function generator), PG (pulse generator), EA (electrical amplifier), ISO: optical isolator. (b) Continuous-wave (CW) Raman seed: SOA (semiconductor optical amplifier), OC (optical coupler), BPF (bandpass filter). (c) SRS fiber amplifier: WDM (wavelength division multiplexer), (d) NIR-PAM system: CL (collimation lens), OL (objective lens), WT (water tank), UST (ultrasound transducer), DAQ (data acquisition system).

Figure 1 shows the schematic diagram of the proposed high spectral energy density all-fiber nanosecond pulsed 1.7 μm SRS amplifier integrated into a custom-built NIR-PAM system. To realize a narrow-linewidth of 1719.44 nm for nanosecond pulsed light source using a CW Raman seed light, we meticulously designed the light source's configuration with consideration of the following key aspects: **(1)** the peak power of the pulsed Raman pump must reach the 1$^{st}$ Raman threshold to ensure sufficient pulse energy in the several microjoule range, **(2)** the Raman seed light should exhibit a narrow-linewidth (~0.10 nm) at the target Stokes wavelength, and **(3)** the frequency spacing between the pulsed Raman pump and the Raman seed light should fall within the range of 11–15 THz [18,21,43,45–47]. Based on these design considerations, we ultimately selected a 1589.80 nm pulsed Raman pump and a 1719.44 nm Raman seed light with a linewidth of 0.09 nm, in order to achieve nanosecond pulses with microjoule-level pulse energy precisely at the lipid absorption peak of 1719.44 nm.

### *2.2. MOFA configured pulsed Raman pump*

The pulsed Raman pump was configured using a MOFA architecture, seeded by a distributed feedback laser diode (DFB-LD). In typical silica fibers, the peak value of the Raman gain spectrum is located around 13.2 THz, with an effective bandwidth ranging of 11–15 THz [18,21,43,45–47]. To generate Stokes light at 1719.44 nm—corresponding to the lipid absorption peak—the pulsed Raman pump wavelength must be carefully positioned within the L-band region to satisfy the required Raman frequency shift of a silica fiber. Taking into account the gain characteristics of erbium-doped and erbium-ytterbium co-doped fiber amplifiers (EDFAs and EYDFAs), we designed the pulsed Raman pump wavelength at 1589.80 nm, which corresponds to a Raman frequency shift of approximately 14.3 THz from target Stokes wavelength, offering an efficient energy transfer from the pulsed Raman pump to the target Stokes wavelength [47]. This pulsed Raman pump wavelength enabled efficient energy transfer to the 1719.44 nm Stokes wavelength, facilitating strong and spectrally aligned SRS amplification.

As shown in Fig. 1(a), the pulsed Raman pump unit consisted of a DFB-LD, an electro-optic modulator (EOM), and a series of optical amplifiers. The DFB-LD was directly modulated with a 10 ns electrical pulse generated by an arbitrary function generator (AFG), which was subsequently modulated by an EOM to achieve a narrower pulse width of 5 ns. The 5 ns pulse was produced using a pulse generator (PG) and amplified by an electrical amplifier (EA). The modulated signal was then amplified through a multi-stage amplifier.

The amplifiers comprised three sequential amplification stages: namely, a pre-amplifier (1$^{st}$-stage EDFA), followed by power-amplifiers (two EYDFAs) serving as the 2$^{nd}$ and 3$^{rd}$ amplification stages. This configuration compensated for the inherently lower gain in the L-band and ensured sufficient pulse energy to exceed the Raman threshold power. However, at higher erbium (Er) ion concentrations—often required for strong amplification—EDFAs are prone to ion clustering, which leads to pair-induced quenching and reduction in stimulated emission efficiency under laser excitation [48]. This phenomenon limits the achievable gain in single-doped fiber amplifiers such as EDFAs. To address this issue, a co-doped architecture employing both Er and ytterbium (Yb) ions was adopted in the later amplification stages. The Yb ions served as sensitizers, facilitating energy transfer and buffering the Er ions, thereby enhancing the overall amplification efficiency [49]. Based on these design considerations, we utilized an EDFA in the 1$^{st}$ pre-amplifier stage, followed by EYDFA modules for both the 2$^{nd}$ and the 3$^{rd}$ power-amplifier stages. Bandpass filters (BPFs) and optical isolators (ISOs) were incorporated between stages to preserve narrow-linewidth and suppress backward reflections. To characterize the spectral and temporal properties of the system without disturbing the main output path, 1% tap port of 99:1 optical couplers (OCs) were placed after the 1st EDFA (pre-amplifier) and 3rd EYDFA (power-amplifier) stages for monitoring.

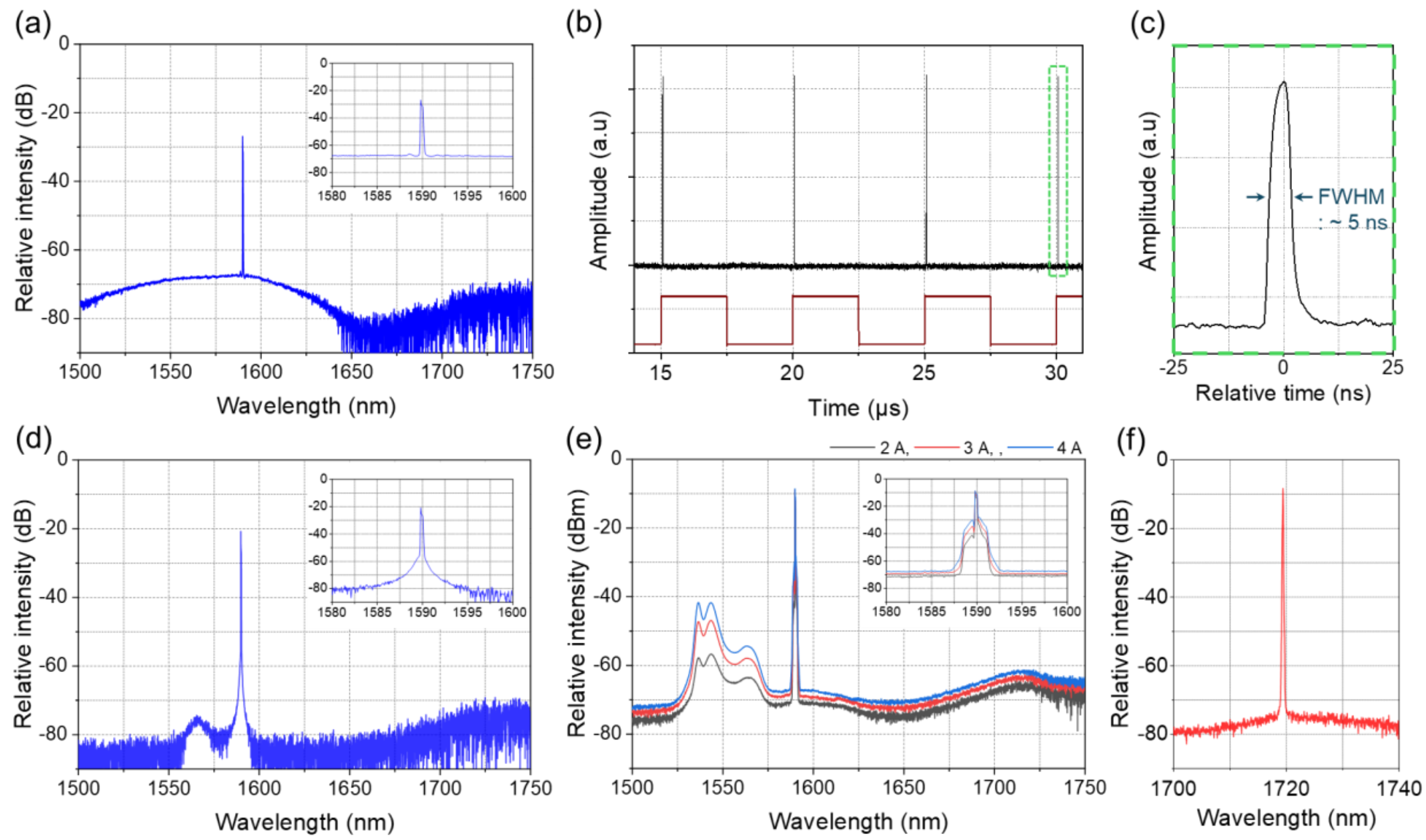

**Fig. 2.** Output characteristics of the pulsed Raman pump and custom-built CW Raman seed laser. (a) Optical spectrum after the EOM [point A in Fig. 1(a)]. (b) Time-trace after the EOM [point A in Fig. 1(a)]. (c) Enlarged view of the pulse temporal profile after the EOM. (d) Optical spectrum after the 1st erbium-doped fiber amplifier (EDFA) stage [point B in Fig.1(a)]. (e) Optical spectrum after the 3rd erbium-ytterbium co-doped fiber amplifiers (EYDFA) stage [point C in Fig.1(a)]. (f) Optical spectrum of the custom-built CW Raman seed laser [point D in Fig.1(c)]. Inset figures in (a), (d), and (e) show enlarged views of the 1580–1600 nm spectral region.

Figure 2(a-e) presents the output characteristics of the pulsed Raman pump measured at various points along the light source, as indicated in Fig. 1. Figure 2(a) shows the optical spectrum immediately following the EOM, centered at 1589.80 nm, while the corresponding temporal pulse profile is illustrated in Fig. 2(b-c). As shown in Fig. 2(c), the full width at half maximum (FWHM)

of the pulses is measured to be approximately 5 ns, resulting from an electrical pulse modulation signal applied to the EOM. Figure 2(d) presents the optical spectrum measured after the 1st EDFA (pre-amplifier) stage, where the average output power was increased to approximately 22 mW with a side-mode-suppression-ratio (SMSR) of approximately 55 dB. Subsequent amplification by the 2nd and 3rd EYDFAs (power-amplifiers) further enhances the signal, as shown in Fig. 2(e). At drive currents of 2.0, 3.0, and 4.0 A, the corresponding average output powers were measured to be 210, 360, and 525 mW, respectively. Through these three amplification stages, the total optical gain reached approximately 37 dB, amplifying the signal from 100 μW to 525 mW while preserving spectral integrity—an essential condition for efficient and stable Raman conversion.

*2.3. Narrow-linewidth CW Raman seed*

In the absence of Raman seed light and/or passive components such as fiber Bragg grating mirror sets, the SRS process in silica fibers is dominated by the intrinsic Raman gain profile centered around a 13.2 THz frequency downshift. This results in a broad Raman gain spectrum spanning of 11–15 THz [43,45], as is typical in standard silica fibers (red line in Fig. 3a). By introducing a narrow-linewidth Raman seed laser—such as a DFB-LD—the SRS output can be spectrally confined, leveraging the selective Raman gain characteristics of the silica fiber [18,46,47,50]. This spectral narrowing is essential for achieving accurate molecular specificity in bond-selective PAM systems. Although narrow-linewidth Raman seed lasers are commercially available for many wavelength regions, certain spectral bands—including the 1700 nm region—remain unsupported by standard off-the-shelf components. To overcome this limitation, we developed a custom-built CW Raman seed laser centered at 1719.44 nm. As shown in Fig. 1(b), it consists of a semiconductor optical amplifier (SOA), a BPF, an ISO, and an OC, collectively ensuring a narrow-linewidth and precise wavelength stabilization at the target Stokes output. As shown in Fig. 2(f), the output spectrum of the CW Raman seed laser is centered at approximately 1719.44 nm, with a measured spectral linewidth of 0.09 nm. The laser exhibits an average output power of approximately 10 mW and an SMSR exceeding 60 dB.

The CW Raman seed laser was implemented using a simple ring cavity configuration, which enables significantly narrower linewidths compared to the use of a BPF alone, whose intrinsic bandwidth is approximately 0.25 nm [51,52]. This architecture enhances wavelength selectivity through mode discrimination within the cavity, reducing amplified spontaneous emission and stabilizing the output spectrum. Moreover, the ring-cavity design offers potential for broadband wavelength tunability—a feature that is typically unachievable with conventional CW Raman seed lasers such as DFB-LDs.

*2.4. SRS fiber amplifier and NIR-PAM system*

Using the pulsed Raman pump and seed lights, the combined light was injected into the specially designed SRS fiber amplifier, as illustrated in Fig. 1(c). The theoretical Raman threshold power, $P_{th}$, for SRS in the gain medium can be described by the following expression:

$$P_{\mathrm{th}} \approx \frac{16A_{eff}}{L_{eff}\, g_R} \tag{1}$$

where $g_R$ is the Raman gain coefficient, $L_{eff}$ is the effective fiber length, and $A_{eff}$ indicates the effective core area. In this study, Raman fibers with a high $g_R$ and optimized $L_{eff}$ were employed to efficiently generate 1$^{st}$ Stokes light with sufficient pulse energies for subsequent amplification [18,22].

Within the SRS fiber amplifier, narrow-linewidth 1$^{st}$ Stokes light is generated through the interaction between the amplified pulsed Raman pump and the injected CW Raman seed light at 1719.44 nm. Compared to conventional all-fiber light sources—such as SC lasers or SPRFLs—the resulting Stokes output exhibits a significantly narrower linewidth, leading to a higher spectral energy density of the pulses.

The output light of the SRS fiber amplifier is delivered to the custom-built NIR-PAM system, as illustrated in Fig. 1(d). The output light is first collimated using a collimation lens (CL, Thorlabs C40APC-C) and then focused onto the target sample through a 10× objective lens (OL, Mitutoyo MY10X-823). The sample, immersed in a water tank (WT), undergoes thermoelastic expansion upon pulsed optical excitation, generating photoacoustic (PA) waves. These waves are detected by a 25 MHz ultrasound transducer (UST, Olympus V324-SU), which converts them into corresponding electrical signals. The theoretical resolution is approximately 70 μm. The signals are filtered through a bandpass filter (BPF, JSR DPR300) and amplified by electrical amplifiers (EAs, Mini-circuits ZFL-500LN+) before being digitized via a data acquisition (DAQ, Alazartech ATS9353) system. The sample was scanned using a two-dimensional motorized translation stage operating in a raster-scanning mode to enable wide-field image acquisition. The motorized stage was synchronized with the NIR-PAM system via a TTL trigger signal from the AFG, which simultaneously controlled the modulation DFB-LD. This configuration ensured precise timing alignment between the proposed light source and the raster-scanning sequence. Raw PA signals acquired from the DAQ were averaged over 10 times to improve the SNR. The PA signals were then followed by envelope detection via Hilbert transform and absolute peak detection. For intuitive visualization, maximum amplitude projection was applied to generate en face images from three-dimensional PA data. The entire NIR-PAM system was controlled via custom-built

LabVIEW software, while ImageJ was used for post-processing, image enhancement, and visualization of the acquired data.

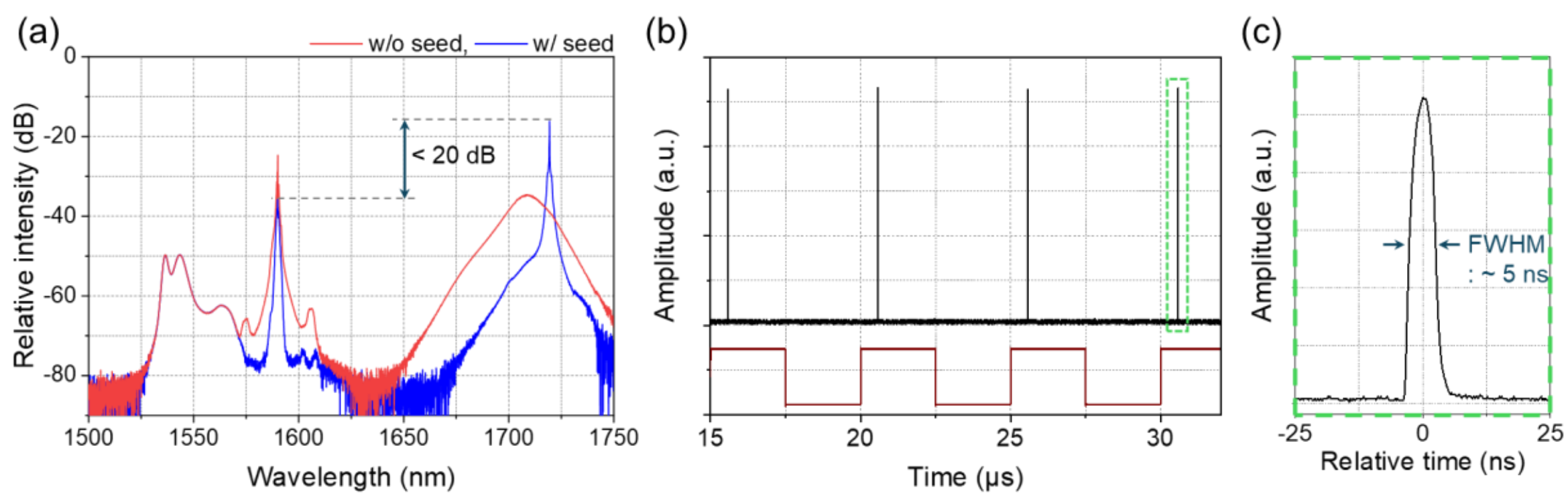

**Fig. 3.** Output characteristics of the proposed SRS fiber amplifier [point E in Fig.1(c)]. (a) Optical spectra with/without the Raman seed light (b) Time-trace with the Raman seed light (c) Enlarged view of the pulse temporal profile.

Figure 3 illustrates the output characteristics of the system following the SRS fiber amplifier. Figure 3(a) compares the output spectra with and without the Raman seed light injection. As expected, in the absence of Raman seed light, a broad 1$^{st}$ Stokes light emission appeared near the 1710 nm region, corresponding to a Raman shift of approximately 13.2 THz. In contrast, when the Raman seed light was injected, a spectrally narrow 1$^{st}$ Stokes light centered at 1719.44 nm (~14.3 THz shift) was generated, exhibiting a linewidth of approximately 0.10 nm. The power ratio between the 1$^{st}$ Stokes light and the residual 1589.80 nm pulsed Raman pump light was measured to be approximately 20 dB, indicating efficient energy transfer through seeded SRS.

Figures 3(b) and (c) present the temporal profile and enlarged time trace of the Stokes output with the Raman seed light, revealing a pulse width of approximately 5 ns, which is consistent with the modulated pulsed Raman pump light shown in Fig. 2(c). The average power of Stokes output was measured to be approximately 440 mW, corresponding to a pulse energy of approximately 2.2 μJ at a PRR of 200 kHz. The spectral energy density of Stokes output was calculated to be approximately 22 μJ/nm, which is particularly remarkable given the narrow-linewidth nature of the proposed light source. This value is substantially higher than those of conventional 1.7 μm light sources—SC lasers and SPRFLs typically yield pulse energy densities of approximately 30 [33] and 500 nJ/nm [13], respectively. The high spectral energy density achieved by the proposed SRS fiber amplifier system—combined with its narrow-linewidth—is crucial for the effective and selective PA excitation of specific molecular bonds, offering a distinct advantage for high-contrast, bond-selective PAI.

# 3. Experimental Results

*3.1. Characterization of the 1.7 μm pulsed SRS fiber amplifier with tunable PRR*

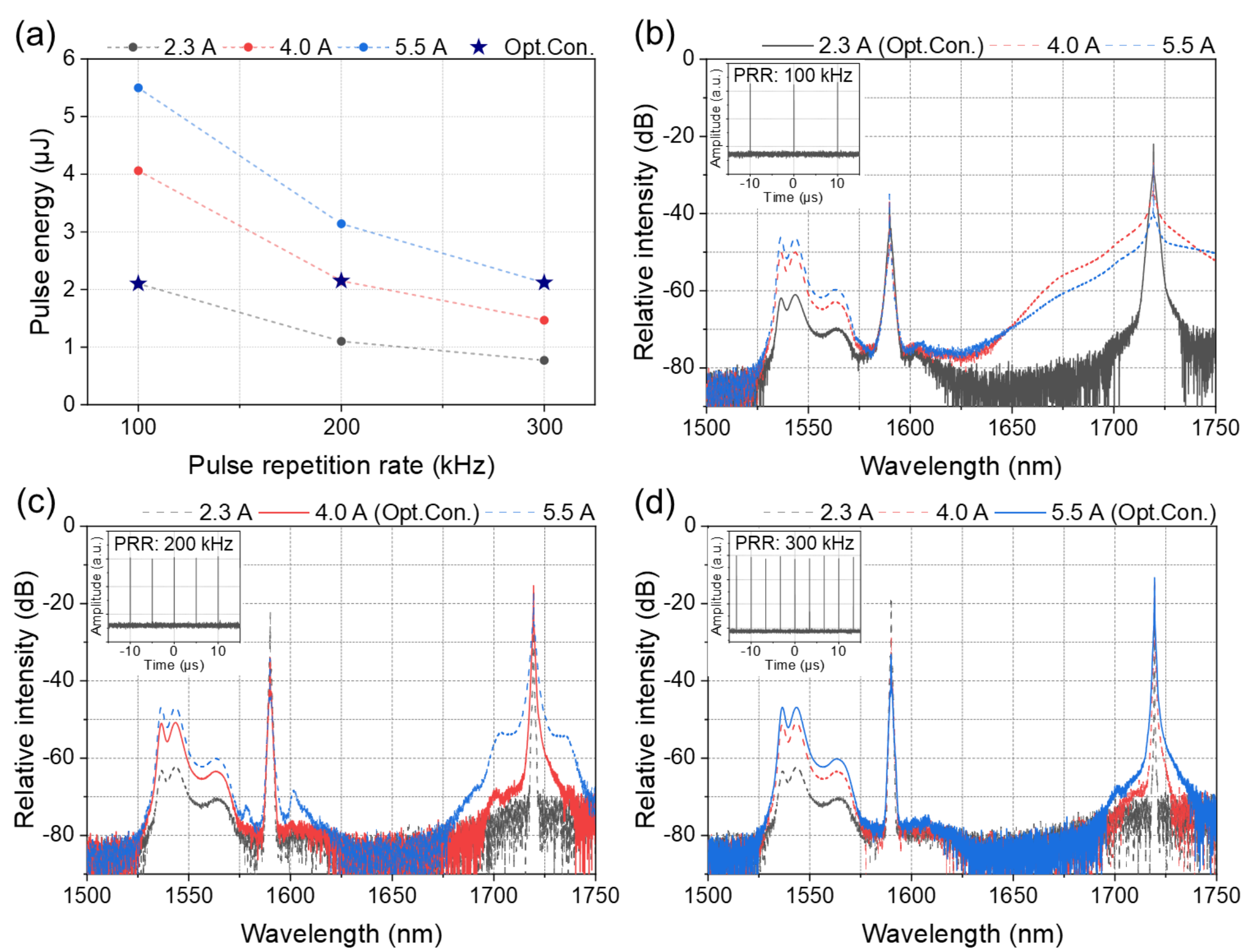

**Fig. 4.** Pulse energy and output spectrum as a function of pulse repetition rate (PRR) and applied current for the 3rd EYDFA. (a) Variation of pulse energy with different PRRs and applied currents. Output spectrum under various applied currents at PRR of (b) 100 kHz, (c) 200 kHz, and (d) 300 kHz, Opt. Con. (optimal condition). Inset figures in (b)–(d) show time-domain pulse trains illustrating the applied PRRs.

One of the key advantages of the proposed SRS fiber amplifier is its PRR tunability, which enables both fast image acquisition and enhanced SNR through temporal averaging—a standard technique in the PAM system. This feature was experimentally evaluated in terms of output pulse energy and spectral response under varying drive conditions, as summarized in Fig. 4. Figure 4(a) shows the variation in pulse energy at three different PRR settings—100, 200, and 300 kHz—corresponding to applied currents of 2.3, 4.0, and 5.5 A for the 3rd EYDFA stage. Across all settings, a pulse energy of approximately 2.2 μJ was achieved under optimized conditions. Figures 4(b–d) compare the output spectra under optimal (solid lines) and non-optimal (dashed lines) conditions. The optimal conditions, defined by the combination of the narrowest linewidth and the highest SMSR, were observed at 2.3 A for 100 kHz, 4.0 A for 200 kHz, and 5.5 A for 300 kHz, respectively. These results confirm that the proposed light source supports flexible PRR adjustment, allowing performance optimization through precise control of the drive current. This flexibility is particularly beneficial in optical imaging systems, where higher PRRs facilitate signal averaging to improve SNR and imaging sensitivity—advantages not afforded by fixed-PRR light sources.

*3.2. Output power stability evaluation*

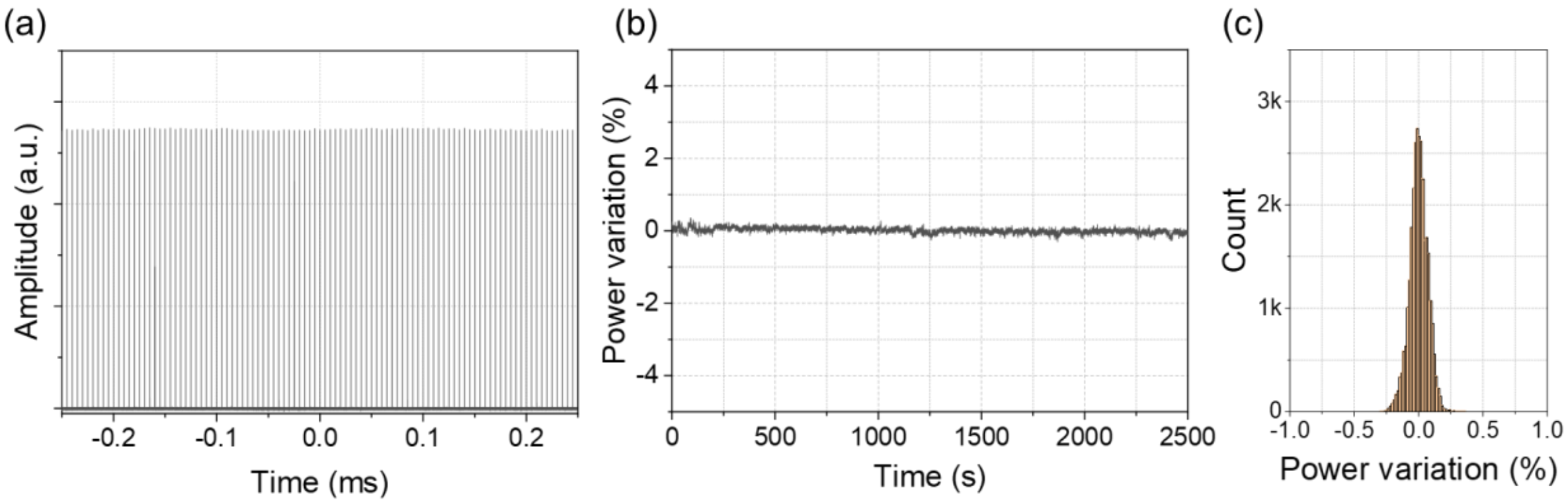

**Fig. 5.** Evaluation of optical output power stability at a PRR of 200 kHz. (a) Short-term output stability measured over a 0.5 ms duration (100 pulses), (b) long-term output stability over a continuous 2500 s duration, and (c) histogram of the long-term power distribution.

Figure 5 presents the results of short-term and long-term power stability measurements under a 200 kHz PRR condition. Since the pulse energy delivered to the target tissue directly affects the amplitude of the PA signal, stable optical output is essential for consistent system performance. Figure 5(a) shows the short-term power stability within a 0.5-ms time window, where 100 consecutive pulses exhibit minimal fluctuation, indicating excellent pulse-to-pulse consistency. Figure 5(b) illustrates the long-term power stability over a continuous acquisition period of 2500-s. This evaluation is particularly important in stage-based scanning systems, which often require long acquisition time. The proposed light source maintains stable output with negligible drift or degradation throughout the entire test duration. Figure 5(c) provides a histogram of the measured power values from the long-term test, revealing a standard deviation (SD) of less than 0.07%, thereby confirming the source's reliability for extended imaging sessions.

*3.3. SNR improvement via averaging*

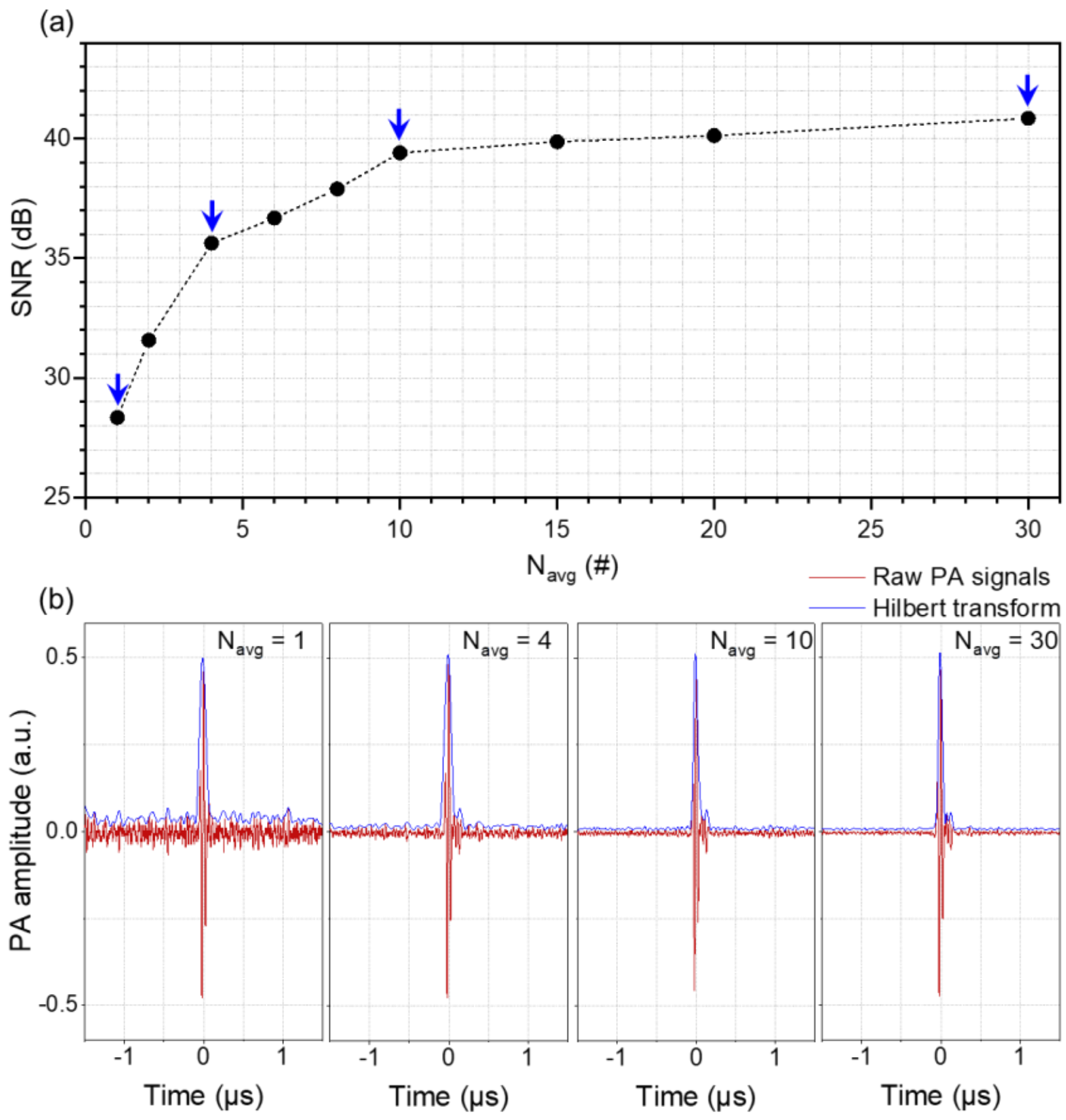

**Fig. 6.** SNR evaluation based on the number of signal averages using a polyethylene (PE) phantom: (a) SNR variation curve as function of averaging number. (b) Representative PA waveforms and corresponding Hilbert-transformed envelopes obtained after 1, 4, 10, and 30 averages, respectively

The proposed system operates at a high PRR, enabling effective signal averaging to enhance the SNR of the PA signal. To evaluate this capability, we analyzed SNR improvements as a function of the number of signal averages. Polyethylene (PE) films were used as the target sample due to their high content of $–CH_2–$ groups [53,54], making them chemically analogous to the lipid molecules targeted in this study. As shown in Fig. 6(a), the SNR increases rapidly from approximately 28 dB at 1 to 39 dB at 10 averages, beyond which the improvement becomes marginal, indicating diminishing returns. While higher number of averages generally lead to better SNR, it also proportionally increases the total data acquisition time. Therefore, selecting an appropriate number of averages is essential to balancing signal quality and acquisition efficiency. In our system, an averaging number of 10 was chosen as the optimal compromise between SNR improvement and imaging speed. Figures 6(b) display representative PA signals obtained after averaging 1, 4, 10, and 30, as indicated by the blue arrows in Fig. 6(a). For intuitive comparison, the Hilbert-transformed signal was displayed in a blue line. Supplementary material includes additional results for other averaging numbers.

### *3.4. Performance of the proposed 1.7 μm NIR-PAM system*

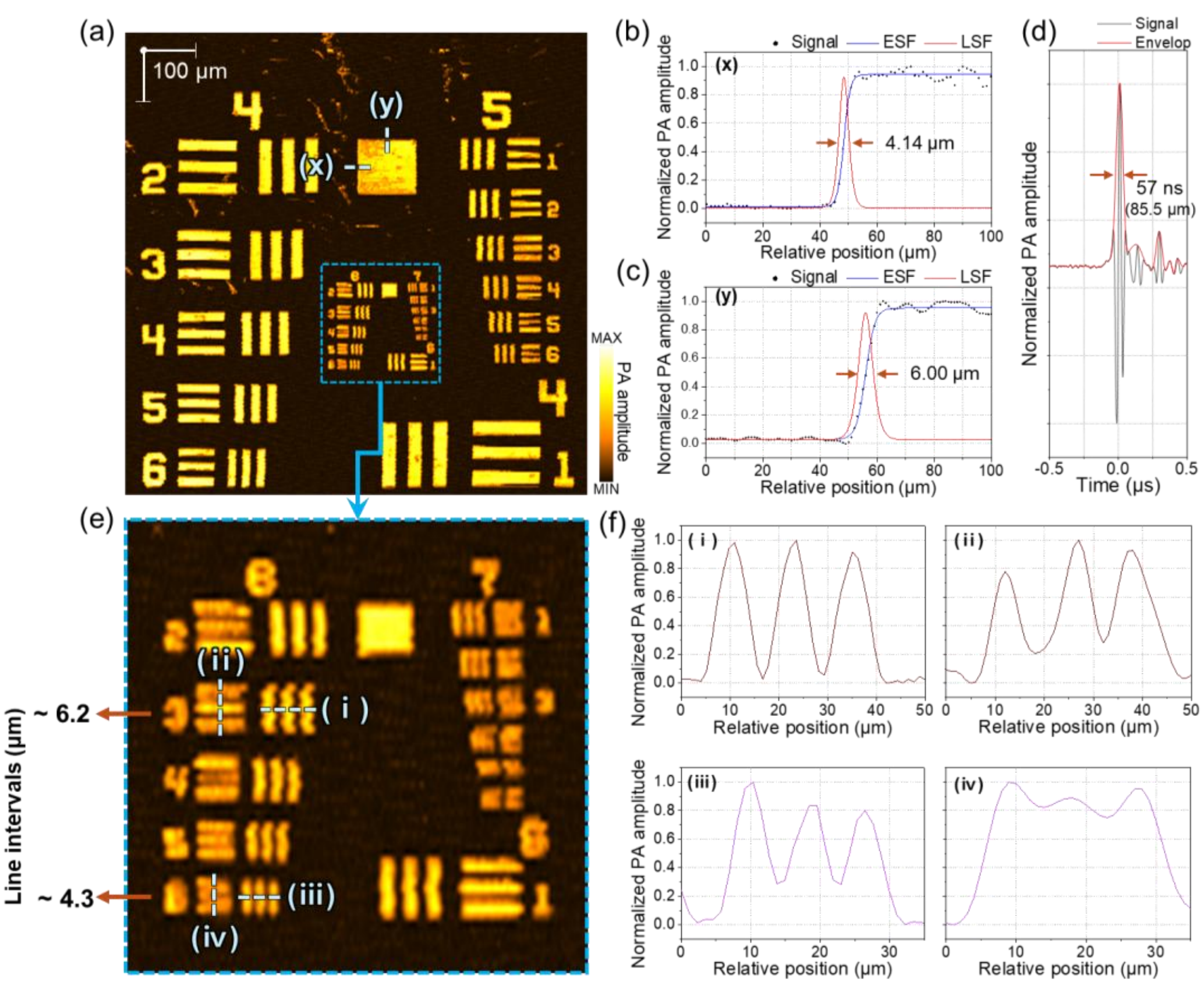

**Fig. 7.** Characterization of the proposed 1.7 μm NIR-PAM system resolution using a 1951 USAF resolution target: (a) PAM image of Groups 4 to 7. (b) Edge profiles and lateral resolution along the x-axis. (c) Edge

profiles and lateral resolution along the y-axis, ESF (edge spread function), LSF (line spread function). (d) Axial resolution extracted from A-line signals. (e) Zoomed-in PAM image of Groups 6 and 7 from Fig. 5(a). (f) Cross-sectional intensity profiles of Group 6 elements—(i) x-axis profiles of Element 3, (ii) y-axis profiles of Element 3, (iii) x-axis profiles of Element 6, and (iv) y-axis profiles of Element 6.

Figure 7 presents the resolution characterization of the proposed 1.7 μm NIR-PAM system, evaluated using a 1951 USAF resolution target. Figure 7(a) displays a wide-field PAM image covering Groups 4 to 7, with a field of view of 0.9 × 0.9 $mm^2$. During the acquisition, step sizes were set to 1.2 μm along the x-axis and 5.0 μm along the y-axis. To ensure uniform pixel scaling, interpolation was applied along the y-axis during post-processing to match the number of data points in both directions. Figures 7(b) and 7(c) show lateral resolution analyses along the x- and y-axes, respectively. In both directions, edge transitions were extracted and fitted to generate edge spread functions, from which the line spread functions (LSFs) were calculated via differentiation. The FWHM of the LSF was used to estimate lateral resolution, yielding values of approximately 4.14 μm along the x-axis and approximately 6.00 μm along the y-axis, which are in good agreement with the theoretical lateral resolution of the NIR-PAM system, calculated to be 4.06 μm. Figure 7(d) displays a representative A-line PA signal used to evaluate axial resolution. Following envelope detection, the FWHM of the detected signal was measured to be approximately 57 ns, corresponding to an axial resolution of approximately 85.5 μm in water. Figure 7(e) provides a zoomed-in view of Groups 6 and 7 for detailed resolution analysis. Figure 7(f) presents cross-sectional intensity profiles for Group 6, Element 3—panels (i) and (ii) correspond to the x- and y-axes profiles, respectively—and for Element 6, panels (iii) and (iv) represent the x- and y-axes profiles. The line spacings in Element 3 and Element 6 are approximately 6.2 and 4.3 μm, respectively. As observed in panels (i), (ii), and (iii), the system clearly resolves three distinct lines, demonstrating its effective lateral resolution. However, in panel (iv), line separation is not discernible due to the system's limited resolution along the y-axis, which results from the larger step size used in that direction. The step size was chosen to balance acquisition time and data volume.

### *3.5. PAM images with the proposed 1.7 μm NIR-PAM system*

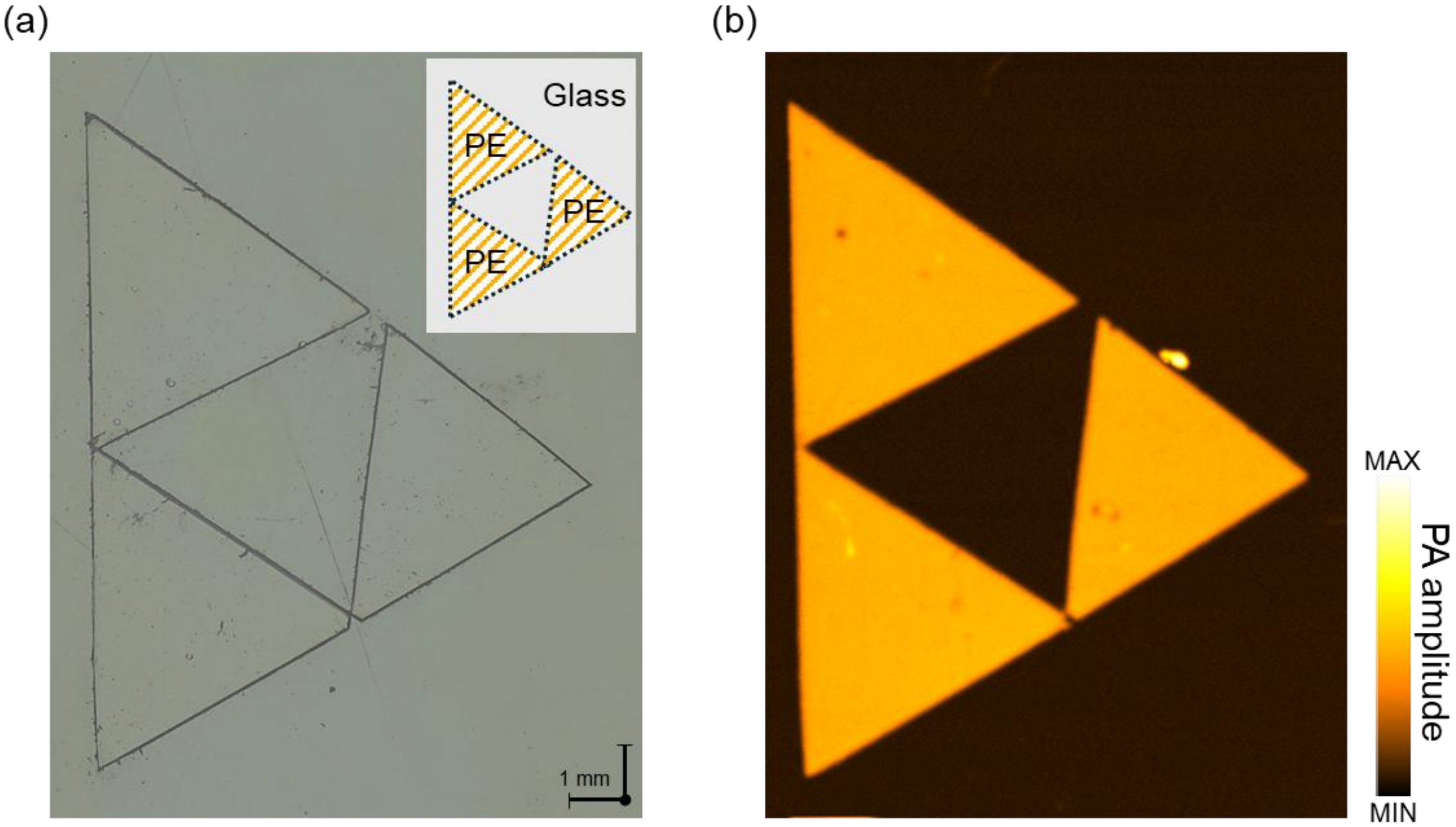

**Fig. 8.** Images of PE films. (a) Stitched 20× optical microscopy image. (b) PAM image.

Figure 8(a) shows a stitched 20× optical microscopy image of the PE films mounted on a glass dish. The sample was fabricated in the shape of a large triangle, featuring smaller triangular patterns etched within to create a structurally complex phantom. The overall configuration of the sample is also shown in the upper-right section. Figure 8(b) presents the corresponding PAM image, which accurately reproduces the geometric features observed in the optical image. The strong PA waves generated from the PE films are attributed to their high concentration of $CH_2$ bonds, which exhibit strong overtone absorption near 1720 nm within the NIR-III window. This absorption closely resembles the spectral characteristics of lipid molecules, making PE an excellent non-biological phantom for evaluating system sensitivity to C–H vibrational modes overtones. These results confirm the system's ability to detect and spatially resolve $CH_2$-rich structures with high fidelity, even within synthetic polymer-based materials.

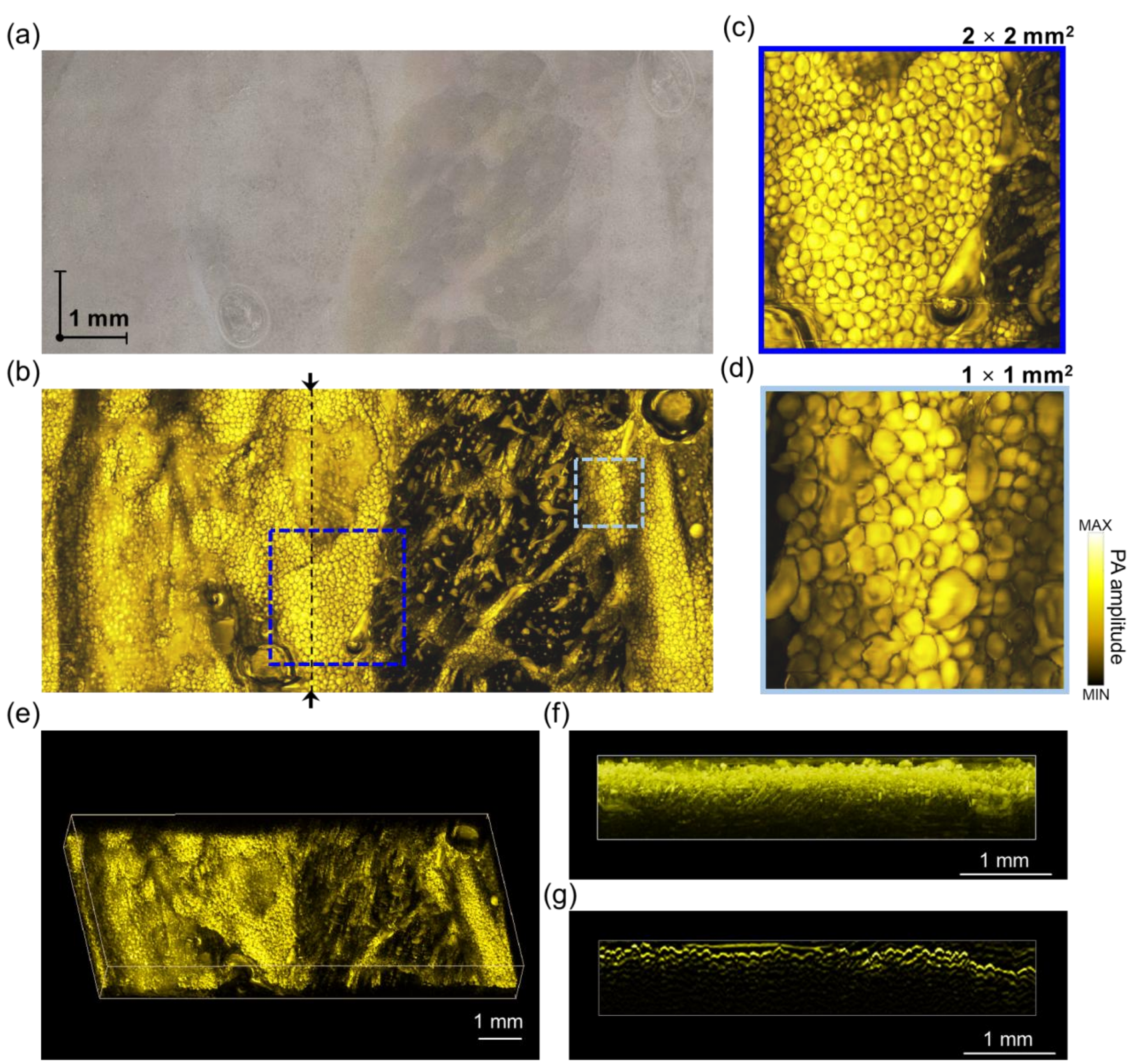

**Fig. 9.** Images of a lipid-rich tissue sample. (a) Stitched 100× optical microscopy image of the lipid-rich tissue sample. (b) Maximum amplitude projection PAM image. Magnified PAM views of selected regions in (b), showing lipid-rich cellular structures over (c) 2 × 2 mm² and (d) 1 × 1 mm² areas, respectively. (e) Oblique Three-dimensional (3D) rendering, (f) side-view, and (g) cross-sectional view of the volumetric PAM image (at black dashed line in Fig. 9(b)).

To demonstrate the applicability of the proposed light source for PAM imaging of endogenous lipid contrast, we conducted experiments on a biologically relevant, lipid-rich tissue sample commonly used in the field. The thickness of the lipid-rich tissue sample was approximately 1 mm. Figure 9(a) shows a stitched optical microscopy image (100× magnification) of the sample, which consisted of a section of pork belly selected for its naturally high lipid content. The bright regions in the optical image correspond to areas with elevated lipid concentrations relative to the surrounding tissue.

Figure 9(b) presents the corresponding maximum amplitude projection PAM image acquired at a PRR of 200 kHz. The scanning step sizes were set to 1 µm along the x-axis and 5 µm along the y-axis. To enhance the SNR, a ten-fold signal averaging was applied along the x-axis. The total acquisition time was approximately 2000 s (~44.4 s/mm$^2$) due to the stage-scanning setup. The resulting PAM image reveals significantly higher PA amplitudes in lipid-rich regions, confirming strong optical absorption of lipids at 1719.44 nm. Zoomed-in views of representative regions in the PAM image are shown in Figs. 9(c) and 9(d), corresponding to 2 × 2 mm² and 1 × 1 mm² areas. These magnified images clearly reveal lipid cell structures, highlighting the system's ability to resolve fine morphological features at the microscale. Figures 9(e-g) display an oblique three-dimensional (3D) rendering, side-view, and a cross-sectional view of the volumetric PAM image, respectively. The cross-sectional line is indicated by a black arrow and dashed line in Fig. 9(b).

These findings collectively validate the effectiveness of the proposed 1.7 μm NIR-PAM system in achieving high-resolution, label-free imaging of lipid distribution in biological tissues—an essential capability for future biomedical diagnostic and research applications.

## 4. Conclusion and Discussion

We developed a high spectral energy density all-fiber nanosecond pulsed 1.7 μm SRS amplifier optimized for NIR-PAM targeting the 1$^{st}$ overtone absorption band of lipids. The light source delivers high pulse energy (≥2.2 μJ), a narrow-linewidth (~0.10 nm), a short pulse duration (~5 ns), and tunable PRR up to 300 kHz. Notably, it achieves a high spectral energy density of approximately 22 μJ/nm, enabling effective and wavelength-selective PA excitation at 1719.44 nm, near the C–H overtone absorption peak in the NIR-III window.

The light source adopted a MOFA architecture, combining flexible PRR control, compact all-fiber construction, and stable spectral output. A 1589.80 nm DFB-LD was selected as the Raman pump laser to achieve an optimal Raman frequency shift of approximately 14.3 THz, effectively aligned with the target wavelength. When paired with a custom-built, narrow-linewidth CW Raman seed laser, the light source delivers highly concentrated excitation energy within a narrow spectral band. The short- and long-term stability of the proposed light source was also evaluated. In particular, the long-term test conducted over 2500 s demonstrated a SD of only 0.07%, confirming its high output stability.

The system's performance was validated through imaging of a 1951 USAF resolution target, achieving a spatial resolution of approximately 4.14 μm, and an axial resolution of approximately 85.5 μm. Phantom experiments using PE films demonstrated the system's ability to resolve structurally complex C–H-based materials. Furthermore, 3D PAM imaging of lipid-rich biological tissue successfully visualized individual lipid cells with high contrast, confirming its utility for high-resolution, label-free molecular imaging.

While the current system demonstrates strong performance, improvements are still needed to fully leverage its high, tunable PRR capability. Imaging acquisition speed is currently limited by mechanical stage scanning. Future work will focus on integrating faster scanning mechanisms—such as micro-electromechanical systems-based or galvanometric mirror scanning—and high-speed signal acquisition and processing pipelines to enable real-time or near real-time imaging. Additionally, we aim to extend the system toward tunable or switchable wavelength operation across the NIR-II and NIR-III regions for multispectral PAI. Achieving this will require the co-design of both the pulse Raman pump and the Raman seed light sources, as the optimal frequency shift and gain bandwidth must be carefully aligned. This spectral flexibility would expand the system's applicability beyond lipid imaging. Furthermore, the system's flexible triggering scheme can also facilitate integration with complementary imaging modalities, such as optical coherence tomography and ultrasound, enabling synchronized multimodal signal acquisition.

Although the current system employs a transmission-mode configuration for evaluation, its compact all-fiber design supports stable, microjoule-scale pulse energy output, making it a promising candidate for future in vivo reflection-mode PA imaging. Potential challenges in reflection-mode implementation—particularly the strong water absorption at 1700 nm—can be addressed by using alternative immersion media such as heavy water [55]. Notably, the system's compact all-fiber design enables stable, microjoule-scale pulse energy output, making it a promising candidate for future in vivo reflection-mode PA imaging applications. The feasibility of deep tissue imaging will be further explored through quantitative assessments of depth-dependent PA responses, using both tissue-mimicking phantoms and in vivo models. Moreover, the impact of spectral linewidth on lipid-selective PA imaging will be investigated through comparative studies, with the aim of quantitatively determining how spectral purity influences molecular specificity and image contrast.

Overall, the proposed light source offers a compact, efficient, and high spectral energy density excitation platform for NIR-PAM. It holds strong potential for translation into in vivo, endoscopic, and future multispectral applications, supporting early-stage detection of lipid-associated diseases such as atherosclerosis and metabolic disorders, and paving the way for next-generation translational PAI platforms, including mesoscopy [56] and endoscopy [57].

## Fundings

This work was supported by the National Research Foundation of Korea (NRF) grant funded by the Korea Government (MSIT) (No. NRF2021R1A5A1032937 & No.RS-2024-00427153) and a grant of the Korea Health Technology R&D Project through the Korea Health Industry Development Institute (KHIDI), funded by the Ministry of Health & Welfare, Republic of Korea (No. HR20C0026).

## Declaration of Competing Interest

The authors declare no conflict of interest.

## Data Availability Statement

The data of this work is available from the corresponding author upon reasonable request.

## References

[1] L. V. Wang, Multiscale photoacoustic microscopy and computed tomography, Nat Photonics 3 (2009) 503–509. https://doi.org/10.1038/nphoton.2009.157.

[2] L. V. Wang, S. Hu, Photoacoustic Tomography: In Vivo Imaging from Organelles to Organs, Science (1979) 335 (2012) 1458–1462. https://doi.org/10.1126/science.1216210.

[3] H.F. Zhang, K. Maslov, G. Stoica, L. V Wang, Functional photoacoustic microscopy for high-resolution and noninvasive in vivo imaging, Nat Biotechnol 24 (2006) 848–851. https://doi.org/10.1038/nbt1220.

[4] S. Jeon, J. Kim, D. Lee, J.W. Baik, C. Kim, Review on practical photoacoustic microscopy, Photoacoustics 15 (2019). https://doi.org/10.1016/j.pacs.2019.100141.

[5] J. Yao, L. V. Wang, Photoacoustic microscopy, Laser Photon Rev 7 (2013) 758–778. https://doi.org/10.1002/lpor.201200060.

[6] A.C. Tam, Applications of photoacoustic sensing techniques, Rev Mod Phys 58 (1986) 381–431. https://doi.org/10.1103/RevModPhys.58.381.

[7] M. Yang, Z. Qu, M. Amjadian, X. Tang, J. Chen, L. Wang, All-fiber three-wavelength laser for functional photoacoustic microscopy, Photoacoustics 42 (2025). https://doi.org/10.1016/j.pacs.2025.100703.

[8] J. Kim, J.Y. Kweon, S. Choi, H. Jeon, M. Sung, R. Gao, C. Liu, C. Kim, Y.J. Ahn, Non-Invasive Photoacoustic Cerebrovascular Monitoring of Early-Stage Ischemic Strokes In Vivo, Advanced Science 12 (2025). https://doi.org/10.1002/advs.202409361.

[9] M. Kim, J.H. Han, J. Ahn, E. Kim, C.H. Bang, C. Kim, J.H. Lee, W. Choi, In vivo 3D photoacoustic and ultrasound analysis of hypopigmented skin lesions: A pilot study, Photoacoustics 43 (2025) 100705. https://doi.org/10.1016/j.pacs.2025.100705.

[10] D. Kim, J. Ahn, D. Kim, J.Y. Kim, S. Yoo, J.H. Lee, P. Ghosh, M.C. Luke, C. Kim, Quantitative volumetric photoacoustic assessment of vasoconstriction by topical corticosteroid application in mice skin, Photoacoustics 40 (2024) 100658. https://doi.org/10.1016/j.pacs.2024.100658.

[11] W. Choi, B. Park, S. Choi, D. Oh, J. Kim, C. Kim, Recent Advances in Contrast-Enhanced Photoacoustic Imaging: Overcoming the Physical and Practical Challenges, Chem Rev 123 (2023) 7379–7419. https://doi.org/10.1021/acs.chemrev.2c00627.

[12] J. Park, S. Choi, F. Knieling, B. Clingman, S. Bohndiek, L. V. Wang, C. Kim, Clinical translation of photoacoustic imaging, Nature Reviews Bioengineering 3 (2024) 193–212. https://doi.org/10.1038/s44222-024-00240-y.

[13] S.W. Cho, T.T.V. Phan, V.T. Nguyen, S.M. Park, H. Lee, J. Oh, C.S. Kim, Efficient label-free in vivo photoacoustic imaging of melanoma cells using a condensed NIR-I spectral window, Photoacoustics 29 (2023). https://doi.org/10.1016/j.pacs.2023.100456.

[14] T. von Knorring, T.B. Ihlemann, P. Blanche, C. Reichl, N.M. Israelsen, C.M. Olesen, Y.T. Yüksel, M. Mogensen, Normal and melanoma skin visualized, quantified and compared by in vivo photoacoustic imaging, Photoacoustics 42 (2025). https://doi.org/10.1016/j.pacs.2025.100693.

[15] T. Feng, Y. Ge, Y. Xie, W. Xie, C. Liu, L. Li, D. Ta, Q. Jiang, Q. Cheng, Detection of collagen by multi-wavelength photoacoustic analysis as a biomarker for bone health assessment, Photoacoustics 24 (2021). https://doi.org/10.1016/j.pacs.2021.100296.

[16] Y. Qiu, H. Li, K. Yu, J. Chen, L. Qi, Y. Zhao, L. Nie, Collagen fibers quantification for liver fibrosis assessment using linear dichroism photoacoustic microscopy, Photoacoustics 42 (2025). https://doi.org/10.1016/j.pacs.2025.100694.

[17] T. Qiu, J. Yang, C. Peng, H. Xiang, L. Huang, W. Ling, Y. Luo, Diagnosis of liver fibrosis and liver function reserve through non-invasive multispectral photoacoustic imaging, Photoacoustics 33 (2023). https://doi.org/10.1016/j.pacs.2023.100562.

[18] H. Lee, M.R. Seeger, N. Lippok, S.K. Nadkarni, G. van Soest, B.E. Bouma, Nanosecond SRS fiber amplifier for label-free near-infrared photoacoustic microscopy of lipids, Photoacoustics 25 (2022). https://doi.org/10.1016/j.pacs.2022.100331.

[19] C. Li, J. Shi, X. Wang, B. Wang, X. Gong, L. Song, K.K.Y. Wong, High-energy all-fiber gain-switched thulium-doped fiber laser for volumetric photoacoustic imaging of lipids, Photonics Res 8 (2020) 160. https://doi.org/10.1364/prj.379882.

[20] J.J.M. Riksen, S. Chandramoorthi, A.F.W. Van der Steen, G. Van Soest, Near-infrared multispectral photoacoustic analysis of lipids and intraplaque hemorrhage in human carotid artery atherosclerosis, Photoacoustics 38 (2024). https://doi.org/10.1016/j.pacs.2024.100636.

[21] S.M. Park, S. Bak, G.H. Kim, C.S. Kim, S.W. Cho, B.E. Bouma, H. Lee, Wavelength-switchable synchronously pumped Raman fiber laser near 1.7 µm for multispectral photoacoustic microscopy, Laser Photon Rev (2024). https://doi.org/10.1002/lpor.202401080.

[22] N.Q. Bui, S.W. Cho, M.S. Moorthy, S.M. Park, Z. Piao, S.Y. Nam, H.W. Kang, C.S. Kim, J. Oh, In vivo photoacoustic monitoring using 700-nm region Raman source for targeting Prussian blue nanoparticles in mouse tumor model, Sci Rep 8 (2018). https://doi.org/10.1038/s41598-018-20139-0.

[23] M. Bondu, C. Brooks, C. Jakobsen, K. Oakes, P.M. Moselund, L. Leick, O. Bang, A. Podoleanu, High energy supercontinuum sources using tapered photonic crystal fibers for multispectral photoacoustic microscopy, J Biomed Opt 21 (2016) 061005. https://doi.org/10.1117/1.jbo.21.6.061005.

[24] S.M. Park, G.H. Kim, H.D. Lee, C.S. Kim, Wavelength-switchable ns-pulsed active mode locking fiber laser for photoacoustic signal generation, Opt Laser Technol 115 (2019) 441–446. https://doi.org/10.1016/j.optlastec.2019.02.045.

[25] S. Zhu, B.C. Yung, S. Chandra, G. Niu, A.L. Antaris, X. Chen, Near-Infrared-II (NIR-II) bioimaging via Off-Peak NIR-I fluorescence emission, Theranostics 8 (2018) 4141–4151. https://doi.org/10.7150/thno.27995.

[26] T.W. Mitchell, H. Pham, M.C. Thomas, S.J. Blanksby, Identification of double bond position in lipids: From GC to OzID, Journal of Chromatography B 877 (2009) 2722–2735. https://doi.org/10.1016/j.jchromb.2009.01.017.

[27] Shabana, S.U. Shahid, S. Sarwar, The abnormal lipid profile in obesity and coronary heart disease (CHD) in Pakistani subjects, Lipids Health Dis 19 (2020). https://doi.org/10.1186/s12944-020-01248-0.

[28] J. Soppert, M. Lehrke, N. Marx, J. Jankowski, H. Noels, Lipoproteins and lipids in cardiovascular disease: from mechanistic insights to therapeutic targeting, Adv Drug Deliv Rev 159 (2020) 4–33. https://doi.org/10.1016/j.addr.2020.07.019.

[29] M. Di Cesare, D.V. McGhie, P. Perel, J. Mwangi, S. Taylor, B. Pervan, C. Kabudula, J. Narula, H. Bixby, D. Pineiro, T.A. Gaziano, F.J. Pinto, The Heart of the World, Glob Heart 19 (2024). https://doi.org/10.5334/gh.1288.

[30] L.P. Dawson, M. Lum, N. Nerleker, S.J. Nicholls, J. Layland, Coronary Atherosclerotic Plaque Regression: JACC State-of-the-Art Review, J Am Coll Cardiol 79 (2022) 66–82. https://doi.org/10.1016/j.jacc.2021.10.035.

[31] H.W. Wang, N. Chai, P. Wang, S. Hu, W. Dou, D. Umulis, L. V. Wang, M. Sturek, R. Lucht, J.X. Cheng, Label-free bond-selective imaging by listening to vibrationally excited molecules, Phys Rev Lett 106 (2011). https://doi.org/10.1103/PhysRevLett.106.238106.

[32] P. Wang, H.W. Wang, M. Sturek, J.X. Cheng, Bond-selective imaging of deep tissue through the optical window between 1600 and 1850 nm, J Biophotonics 5 (2012) 25–32. https://doi.org/10.1002/jbio.201100102.

[33] P. Wang, J.R. Rajian, J.X. Cheng, Spectroscopic imaging of deep tissue through photoacoustic detection of molecular vibration, Journal of Physical Chemistry Letters 4 (2013) 2177–2185. https://doi.org/10.1021/jz400559a.

[34] ANSI.2012: American National Standard for Safe Use of Lasers in Research, Development, or Testing, ANSI Z136.8-2012, American National Standards Institute, Washington, DC, 2012.

[35] L.A. Sordillo, Y. Pu, S. Pratavieira, Y. Budansky, R.R. Alfano, Deep optical imaging of tissue using the second and third near-infrared spectral windows, J Biomed Opt 19 (2014) 056004. https://doi.org/10.1117/1.jbo.19.5.056004.

[36] P.K. Upputuri, M. Pramanik, Photoacoustic imaging in the second near-infrared window: a review, J Biomed Opt 24 (2019) 1. https://doi.org/10.1117/1.jbo.24.4.040901.

[37] T. Zhao, A.E. Desjardins, S. Ourselin, T. Vercauteren, W. Xia, Minimally invasive photoacoustic imaging: Current status and future perspectives, Photoacoustics 16 (2019). https://doi.org/10.1016/j.pacs.2019.100146.

[38] T. Buma, N.C. Conley, S.W. Choi, Multispectral photoacoustic microscopy of lipids using a pulsed supercontinuum laser, Biomed Opt Express 9 (2018) 276. https://doi.org/10.1364/boe.9.000276.

[39] S.W. Cho, S.M. Park, B. Park, D.Y. Kim, T.G. Lee, B.M. Kim, C. Kim, J. Kim, S.W. Lee, C.S. Kim, High-speed photoacoustic microscopy: A review dedicated on light sources, Photoacoustics 24 (2021). https://doi.org/10.1016/j.pacs.2021.100291.

[40] L. Jin, Y. Liang, Fiber laser technologies for photoacoustic microscopy, Vis Comput Ind Biomed Art 4 (2021) 11. https://doi.org/10.1186/s42492-021-00076-y.

[41] H. Al-Taiy, N. Wenzel, S. Preußler, J. Klinger, T. Schneider, Ultra-narrow linewidth, stable and tunable laser source for optical communication systems and spectroscopy, Opt Lett 39 (2014) 5826. https://doi.org/10.1364/OL.39.005826.

[42] Y. Song, S.M. Park, Y. Jeong, J. Kim, H. Lee, Review on Multispectral Photoacoustic Imaging Using Stimulated Raman Scattering Light Sources, Sensors 25 (2025) 3325. https://doi.org/10.3390/s25113325.

[43] G.P. Agrawal, Fiber Optic Raman Amplifiers, in: Guided Wave Optical Components and Devices, Elsevier, 2006: pp. 131–153. https://doi.org/10.1016/B978-012088481-0/50010-3.

[44] H. Lee, G.H. Kim, M. Villiger, H. Jang, B.E. Bouma, C.-S. Kim, Linear-in-wavenumber actively-mode-locked wavelength-swept laser, Opt Lett 45 (2020) 5327. https://doi.org/10.1364/ol.397715.

[45] R.H. Stolen, E.P. Ippen, Raman gain in glass optical waveguides, Appl Phys Lett 22 (1973) 276–278. https://doi.org/10.1063/1.1654637.

[46] H. Lee, M.R. Seeger, B.E. Bouma, Electronically Controlled Dual-Wavelength Switchable SRS Fiber Amplifier in the NIR-II Region for Multispectral Photoacoustic Microscopy, Laser Photon Rev (2024). https://doi.org/10.1002/lpor.202400144.

[47] M. Eibl, S. Karpf, H. Hakert, T. Blömker, J.P. Kolb, C. Jirauschek, R. Huber, Pulse-to-pulse wavelength switching of a nanosecond fiber laser by four-wave mixing seeded stimulated Raman amplification, Opt Lett 42 (2017) 4406. https://doi.org/10.1364/ol.42.004406.

[48] E. Desurvire, M.N. Zervas, Erbium-Doped Fiber Amplifiers: Principles and Applications, Phys Today 48 (1995) 56–58. https://doi.org/10.1063/1.2807915.

[49] H.M. Obaid, H. Shahid, Numerical achievement of high and flat gain using Er-Yb co-doped fiber/Raman hybrid optical amplifier, Optik (Stuttg) 186 (2019) 72–83. https://doi.org/10.1016/j.ijleo.2019.04.089.

[50] T. Sylvestre, H. Maillotte, E. Lantz, P.T. Dinda, Raman-assisted parametric frequency conversion in a normally dispersive single-mode fiber, 1999. https://doi.org/10.1364/OA_License_v1#VOR.

[51] L. Huang, C. Yang, T. Tan, W. Lin, Z. Zhang, K. Zhou, Q. Zhao, X. Teng, S. Xu, Z. Yang, Sub-kHz-Linewidth Wavelength-Tunable Single-Frequency Ring-Cavity Fiber Laser for C- And L-Band Operation, Journal of Lightwave Technology 39 (2021) 4794–4799. https://doi.org/10.1109/JLT.2021.3074824.

[52] Deyu Zhou, P.R. Prucnal, I. Glesk, A widely tunable narrow linewidth semiconductor fiber ring laser, IEEE Photonics Technology Letters 10 (1998) 781–783. https://doi.org/10.1109/68.681482.

[53] M. Tasumi, S. Krimm, Crystal vibrations of polyethylene, J Chem Phys 46 (1967) 755–766. https://doi.org/10.1063/1.1840736.

[54] B. Kanyathare, B.O. Asamoah, U. Ishaq, J. Amoani, J. Räty, K.E. Peiponen, Optical transmission spectra study in visible and near-infrared spectral range for identification of rough transparent plastics in aquatic environments, Chemosphere 248 (2020). https://doi.org/10.1016/j.chemosphere.2020.126071.

[55] C.M. Salinas, E. Reichel, A. Gupta, R.S. Witte, Heavy water coupling gel for short-wave infrared photoacoustic imaging, J Biomed Opt 28 (2023). https://doi.org/10.1117/1.JBO.28.11.116001.

[56] H. He, C. Fischer, U. Darsow, J. Aguirre, V. Ntziachristos, Quality control in clinical raster-scan optoacoustic mesoscopy, Photoacoustics 35 (2024). https://doi.org/10.1016/j.pacs.2023.100582.

[57] Q. Xia, S. Lv, H. Xu, X. Wang, Z. Xie, R. Lin, J. Zhang, C. Shu, Z. Chen, X. Gong, Non-invasive evaluation of endometrial microvessels via in vivo intrauterine photoacoustic endoscopy, Photoacoustics 36 (2024). https://doi.org/10.1016/j.pacs.2024.100589.